\documentclass[manuscript]{aastex}
\usepackage{latexsym,epsfig,graphics,graphicx,fullpage,amsmath}

\begin{document}

\title{Probable detection of HI at $z\simeq 1.3$ from DEEP2 galaxies using the GMRT}
\author{Shiv K. Sethi\altaffilmark{1}, K. S. Dwarakanath\altaffilmark{1}, Chandrashekar Murugesan\altaffilmark{1}}
\altaffiltext{1}{Raman Research Institute, Bangalore 560080, India}

\begin{abstract}
We have observed the DEEP2 galaxies using the Giant Meterwave Radio
Telescope in the frequency band of 610 MHz. There are $\simeq  400$ galaxies in
the redshift range $1.24 < z < 1.36$ and within the field of view
 $\simeq 44'$,
of the GMRT dishes. We have coadded the HI 21 cm-line emissions at the
locations of these DEEP2 galaxies.  We apply stacking on three
different data cubes: primary beam uncorrected, primary beam corrected (uniform weighing ) and primary beam corrected (optimal weighing). 
 We obtain a peak signal strength in the range $8\hbox{--}25 \, \rm 
\mu$Jy/beam for a velocity width in the 
range  $270\hbox{--} 810 \, \rm km \, sec^{-1}$. 
The  error on the signal, computed by bootstrapping, lies in the 
range  $2.5\hbox{--}6 \, \rm \mu$Jy/beam, implying a 2.5--4.7-$\sigma$
 detection of
the signal at $z \simeq 1.3$. 
We compare our results  with  N-body simulations of the
 signal at $z\simeq 1$ and find reasonable agreement. 
We also discuss  the impact of residual continuum and systematics. 
\end{abstract}


\section{Introduction}

To determine the evolution of the HI content of galaxies  as a function of 
redshift is a very important
 input 
into the understanding of the history of gas content and star formation
 in the universe. While 
the study of 
damped Lyman-alpha clouds in absorption gives the evolution of the
 aggregate HI content in the redshift range $0.5 \le z \le 5$ (Prochaska et~al. 2005; Rao et~al. 2006; Noterdaeme et~al. 2009, 2012), the 
determination of HI content of a halo of a given mass remains elusive
 at high redshifts owing
to the faintness of individual halos in the HI 21~cm line emission, e.g. 
  the detection of  $\simeq 10^{10} \, \rm M_\odot$ of 
HI at z$ = 1.3$ would take 400 hours of observation
with the Giant Meterwave Radio Telescope (GMRT) (Bagla et~al. 2010). Direct observation  of HI 21-cm line emission and its detailed modelling 
has only been possible at  $z \simeq 0$ (Zwaan et~al. 2005). For 
selected fields, the HI has been detected in emission for $z \la 0.2$  (Verheijen et~al. 2007; Verheijen et al. 2010; Catinella et~al. 2008).

One possible approach to detect HI in emission at high redshifts 
 rests on the  detection of the fluctuation in the 
redshifted HI emission from high redshifts (Bharadwaj \& Sethi 2001; Chang et~al. 2008, Bharadwaj et~al. 2009). 
Recently the first detection of HI at $z\simeq 0.7$ was reported based on 
cross-correlating the 
density field of optically-selected galaxies  with known redshifts from 
the DEEP2 survey with the
 redshifted 
HI emission (Chang et~al. 2010).  This 
cross-correlation is  non-zero only if  HI is associated   with galaxies 
observed by the DEEP2 survey;  or  this method 
detects the component of HI correlated  with the DEEP2 galaxies. More recently,
a  detection at $z \simeq 0.8$ was reported based on the  cross-correlation
 with WiggleZ galaxies  (Masui et~al. 2012). 
Stacking HI spectra of galaxies at known redshifts, akin to this approach, 
has also been attempted for both field and cluster galaxies at low redshifts
(Lah et~al. 2009; Lah et~al. 2007).

Detailed N-body simulations (Khandai et~al. 2011) showed
 that results of Chang et~al (2010) can be 
consistently reproduced within the framework of density evolution in the 
standard $\Lambda$CDM model. These simulations could estimate the
 possible range of bias and stochasticity of the HI density field, which
are degenerate with the total HI content in the observation of 
Chang et~al. (2010), and therefore 
directly estimate the HI content of the universe at $z \simeq 1$. 
 These  simulations also showed  that stacking of 
HI 21 cm-line emission
 from galaxies of 
known redshifts can lead to a detailed reconstruction of the properties
 of HI halos
at $z \simeq 1$. It was argued that
 GMRT could play an important role in this study both owing
 to its frequency  coverage as well as its field of view.

In this letter  we report GMRT observations of one of the DEEP2 fields and attempt
to estimate the HI content of the universe at $z \simeq 1.3$ using stacking 
of the spectra at the locations of the DEEP2 galaxies.

 Throughout this paper we use the spatially flat (k =0) $\Lambda$CDM model
with  $\Omega_m = 0.24$  and $h = 0.73$. 

\section{Observations and Data Analysis}

DEEP2 redshift survey has made spectroscopic measurements of  
 redshifts of over 25000 
objects in the redshift range $0.7 < z < 1.5$, in four fields of 
approximate angular sizes 2$^{o} \times$ 0.5$^{o}$. 
Based on the examinations of the NVSS images of the the four DEEP2 Fields,
the field centered at 1652+3455 (selected as $'$Zone 
of very low extinction$'$ for DEEP2 observations) was selected for
observations with the GMRT at 610 MHz. Amongst the four fields,
this field contains the minimum number of bright radio sources
within the primary
beams (full width at half maximum $\sim$ 44$'$) of the GMRT antennas at 610 MHz. 
This will ensure that the images
are least affected by dynamic range limitations of GMRT. In any case, dynamic range limitations
will be minimal since this is a spectral line observation. 
This field, which is $2^{\circ} \times 0.5^{\circ}$ needs two  GMRT pointings
at 610~MHz. 
The desired velocity resolution is $\simeq 30 \, \rm  km \, s^{-1}$ to 
detect signals with an expected
width $\simeq 200 \, \rm km \,  s^{-1}$. This velocity width takes into account
both the error in the redshift estimates of the DEEP2 galaxies and the
expected velocity widths of the HI 21 cm-line signals. 
The velocity width of $\sim$ 30 km s$^{-1}$ corresponds to a channel
width of $\sim$  60 KHz at 610 MHz. 

Based on the distribution of DEEP2 galaxies in Right Ascension and Declination
and in the redshift space, and taking into account the optimum range
of frequencies for observations in the 610 MHz band, a frequency range of
601 to 633 MHz was selected. This frequency range corresponds to 
DEEP2 sources in the redshift
range $1.24 < z < 1.36$, distributed over the range of right ascensions
$251^\circ.6 < \alpha <  253^\circ.2$, and declinations $\hbox{+}34^\circ.65 < \delta <  +35^\circ.2$. This region of 
79$^{'} \times$ 33$^{'}$ was covered in
two GMRT pointings, the primary beams (FWHM) of the GMRT dishes
being $\sim$ 44$^{'}$. Each of these pointings were observed for a total of
12 hr with a bandwidth of 32 MHz and 512 channels. This setting gives a
velocity resolution of $\sim$ 31 km s$^{-1}$. One of the two 12 hour
pointings was completely analysed and is discussed below. 

Corrupted data due to non-working antennas and baselines and due to 
the radio frequency interference were flagged using different tasks
in the Astronomical Image Processing Software (AIPS). The data quality
was good with less than 10 \% of the data being flagged. 
A multi-channel continuum data set (32 channels) was formed from 
the spectral line data (512 channels) by averaging
over 16~channels. This continuum data set was used
to produce the best continuum images of this field. The continuum
images so obtained had an RMS value of 85 $\mu$Jy/beam for a 
circular synthesised beam of 10$''$. This image is dynamic
range limited at a value $\sim$ 3200, as is usually the case with most 
 GMRT continuum images.

The gain (complex) solutions obtained above were transferred to the spectral
line data set. The continuum in the spectral line data was subtracted
 in two steps. First, the continuum source
components that made the best continuum images were subtracted from the
spectral line data set (task UVSUB in AIPS). Second, a second-order
base line across frequencies was fit to the resulting spectral line 
data set and was subtracted from the spectral line data set
(task UVLSF in AIPS). This  procedure yielded a
continuum-free data set which was analysed for spectral features.
A spectral line image cube (512 channels) with a spatial resolution of
10$^{''}$ and a spectral resolution of 30.7 km s$^{-1}$ was produced.
This spectral line image cube has an RMS of 260 $\mu$Jy/beam/channel, close to
the expected value of $\sim$ 200 $\mu$Jy/beam/channel.

\section{Results}

From the image cube, we extract spectra at  the positions  of DEEP2 sources
in the field of view. We then shift and add the spectra centered at  the 
respective 
redshifts of the DEEP2 sources. In this procedure, we retain 50 channels on 
either side of the center of addition. If a source lies within 50 channels
from  the edges of the cube, it is excluded from the analysis. This process
retains  489 sources out of a total of 539  for the entire field of view.
We also apply primary  beam correction to the image cube. This process
increases the noise and the  source flux away from the 
phase center.  For the primary beam corrected cube, 
we retain sources only within 30\% of the  primary beam.  There are 
389 sources in this angular area and the  procedure  of coadding yields
371 sources.

We present results from both the primary-beam uncorrected and the primary-beam
 corrected cubes.  To take into account this variation in 
the noise properties we coadd  signal with both uniform and 
optimal weighting.  In optimal weighting,
the spectra are coadded with weights inversely proportional to the square
of the noise in each spectra.  The uniform weighting gives us an unbiased
estimate of the signal (average peak flux of the halos) and allow us to 
compare the flux with theoretical estimates. The optimal weighting maximizes
the signal-to-noise of the detection but constitutes a biased estimate 
of the signal.  The estimated signal strength in the three cases can be written as: 
\begin{eqnarray}
S_{\rm uncorr} & = & {\sum_i S_i \over N} \nonumber \\
S_{\rm corr} & = & {\sum_i S_i/w_i \over N} \nonumber \\
S_{\rm opt} & = & {\sum_i S_i/(w_i \sigma_i^2) \over \sum_i 1/\sigma_i^2}
\end{eqnarray}
Here $S_i$ is the signal at the position of DEEP2 galaxies; $\sigma_i$ is the 
noise in the spectrum at the position of DEEP2 galaxies, and $w_i$ corresponds
to the primary beam correction at the location of the galaxy. 

The primary beam uncorrected analysis allows us to study the noise characteristics of the cube, while the 
corrected cases  gives us a realistic estimate of the strength of the
 spectral line.

In Figure~1,  we show the coadded raw  spectrum  and the  7-channel 
running average spectrum, from the primary beam uncorrected and 
the primary beam corrected image cubes.

\subsection{The signal strength}

In this paper we aim to measure the line flux and its velocity width
of the stacked HI signal.
In a detailed  N-body simulation, Khandai et al. (2011) constructed 
several models of HI distribution at $z \simeq 1$ and 
computed the  stacked spectra.  All these models were  normalized 
to the cumulative HI content  $\Omega_{\rm HI}$ as estimated 
by absorption studies, but  were based on a range of   HI  mass functions, 
and therefore differed in emission properties of the signal.
 The  stacked spectra obtained from these models show
a peak flux between 10--30$\mu$Jy and velocity widths in the range
500-1000$\rm km \, sec^{-1}$.

To extract the relevant information from the coadded signal, we average
the signal with three velocity widths. 
Figure~2 shows the coadded spectra averaged over 9 points  
every 10th channel. Each of the nine
points on the red (dashed) curve in the Figure 
 corresponds to signal added for
$(14+9i) \pm 4$ for $i$ running from   0 and 8; the velocity 
width of the nine point averaging is $270 \, \rm km \, sec^{-1}$.
 Figure~3 
and~4 show the signal averaged over 19~channels ($(12+19i)\pm9$ 
for i from 0 to 4;  velocity width
$570 \, \rm km \, sec^{-1}$) and 29~channels ($(21+29i)\pm14$, velocity width
$770 \, \rm km \, sec^{-1}$), respectively.  The choice of these three
velocity averaging schemes   allows us to estimate the
 velocity width of the central enhancement
and  also gauge 
the contribution of residual 
continuum and systematics on the wings of the line. 

As noted above we present analysis of  three different cubes  of estimating the
coadded signal (Eq~1). The interpretation of the signal in the three 
cases needs to be discussed. 
In the 
primary beam uncorrected case, the signal is dominated by halos close to 
the primary beam 
phase center. In the primary beam corrected uniform-weighting all the
 halos  receive 
equal weight. In the optimal-weighing case, the signal is dominated
 by halos closer 
to the primary beam phase center as the halo flux is weighted as the square of the inverse of the noise
which increases away from the phase center when the primary beam correction is made. This also 
means that one generically expects  the signal in this case to be closer to the estimate of 
the primary beam uncorrected case, as we also find. It should be noted
that the optimal case also give us an unbiased estimate of the 
signal if the HI signal from all the haloes is the same.  

The signal strengths obtained for varying  velocity widths using  
three different data cubes are given in Table~1. 

\subsection{Estimate of error on the signal}

We adopt  three different methods for estimating the error on the
signal strength. 
\begin{itemize}
\item[(a)] The most obvious   method to make an estimate of the 
error on the signal is to estimate the RMS of the coadded signal (Figure~1)
after appropriate velocity averaging (Table~1). 
 While this is the simplest
approach to computing the error on the signal, it has  severe  limitations. 
For a nine-point average, we are 
 left with only nine points to compute the signal and the noise; we 
do not quote this number for the other two velocity averaged spectra.

\item[(b)] We scramble the redshifts of DEEP2 galaxies and adopt 
the same procedure for coadding the spectra as outlined above (9, 19, 29-point
averaged, red, dashed curve in Figure~2, 3, and~4).

 This procedure randomizes galaxy positions (in redshift space) 
 and therefore should 
 yield a null result; it can
also act to suppress  systematics in the original coadded spectra.  
We use this
bootstrapping procedure to obtain 
180--200 different realizations  of the spectrum  (20, 40, 60  realizations 
of  9-, 19- and 29- channel averaged spectra, each   containing 9, 5, or 3 points), which 
 enables us to obtain an estimate of the error in the 
signal. We check for convergence in each case. 

For instance, in Figure~2, we show the average and RMS for 9~spectral 
 point for 20 realizations obtained by scrambling DEEP2 galaxy redshifts.
As all  these  spectral points are equivalent 
we could use all the 180 points to  compute  the error on the signal
strength.  Using  all the 180 spectral points, we obtain an 
 RMS of  (4.3,10, 4.1)~$\mu$Jy  
from these spectra for the 
 primary beam uncorrected and corrected (uniform and optimal)  cubes, respectively (Table~1). Similar procedure is applicable for the other
two velocity widths and  the relevant 
 results for these cases  are shown in  Figure~3 and ~4 and Table~1. 

The bootstrapping procedure described above acts to redistribute
line flux across the entire data and while this suppresses 
the central enhancement it could also create a small line-flux bias
which results in non-zero average for the coadded spectrum. The strength of this bias can 
be computed: as noted above we do not source lying within 
 50 channels on either side of 
spectrum, which leaves us with 412 spectral channels. We randomize the 
 redshift of the object for each spectra and follow the 
coadding process. The number of sources
is $\simeq 400$ and  the average line flux is $\simeq 20\, \mu$Jy with 
a velocity width corresponding  to 20 channels (e.g. Figure~3). This flux 
is redistributed across $\simeq 412\times 400$ channels, which results 
in the maximum  line-flux bias  of $\simeq  1\, \mu$Jy for the entire cube. 

 The probability of the scrambled signal to exceed  the central
 enhancement  is another
indicator of the significance of the detection.
 However, when the 
number of  realizations exceed the total number of data points the 
spectral channels get correlated and  we do not
get independent information of this significance. The total number of 
data points is  the number of velocity-averaged spectral points
for one spectrum multiplied by the the number of sources, which gives
roughly 10000, 5000, and 3000 data points for the 9-, 19- and 29-channel
averaging. When the number of realizations exceed these numbers we 
get spectral correlation, which prevents a statistical 
interpretation of our results. 
 This also means that this procedure cannot give more than
a 3.5$\sigma$ detection (for a normal distribution). 

This procedure gives us another measure  of the significance of 
the signal. We run a large number of realizations (over 1000) to 
understand the nature of the signal in each case. Our findings are 
in broad agreement with the results shown in Table~1: (a) an increase in 
signal-to-noise for optimal weighting (we did not find a single instance
in which the bootstrapped signal exceeded the central enhancement in such
cases), (b) the signal-to-noise is maximum for 19-channel averaging case, (c)
for cases where the signal-to-noise is expected to be smaller than
3$\sigma$,  e.g. primary
beam corrected, 29-channel averaged case, the results from this procedure
are in agreement with Table~1.

\item[(c)] The methods (a) and (b)  do not 
 take into account the noise properties 
 of the whole data cube, which could be affected by other systematics. 
To determine the level of  homogeneity of noise  along lines of sights that
do not contain DEEP2 sources (Blank positions), we use the
data along 40000 lines of sights towards these blank positions.
These blank positions are randomly chosen and are 
separated from the positions of DEEP2 sources by more than a few   synthesized
beams.  These spectra   can be used to obtain
many realizations of coadded spectra for the required number of sources. 
We apply exactly the same coaddition procedure on a set containing 
500 such spectra as for the spectra containing the DEEP2 galaxies. 
We thus obtain  80 realizations of such coadded data sets towards blank 
positions, which are shown in Figure~5.

The average RMS of these raw  spectra is $\simeq 12.6 \, \rm \mu$Jy; the 
error on this  RMS  is  $\simeq 0.8\, \rm \mu$Jy. The small error
 on the RMS  shows that the noise properties of the 
data cube are  stable and robust.

 The average RMS for 
the 9,  19, and  29~channel averaged spectra is  $\simeq \{4.4, 3, 2.3\}
 \, \rm \mu$Jy, respectively (Table~1). 

Information from blank positions can be used to estimate the level of 
significance of the detection in the direction of DEEP2 galaxies. We estimate
the probability for the  signal (for 9-, 19-, or 29-channel 
averaging)  in the blank positions to exceed
the central enhancement detected in the direction of DEEP2 galaxies. 
We find this probability to lie between 0.01--0.03 for the three
cases of channel averaging. It is lowest for the 29-channel average and 
is comparable for the other two cases. The estimated probability 
correspond to between 2-2.5$\sigma$ detection of the signal 
for the primary beam uncorrected
cube.

 This procedure can not be used for the 
primary beam corrected cube as the noise increases away from the pointing center
and therefore is not homogeneous across the cube.

\end{itemize}

\subsection{Total line flux, residual continuum,  and systematics}

The results are displayed in Table~1 and Figure~2, 3,~4 
 can be briefly summarized as: (a) The signal strength is  
almost the same  for both 9 and 19 channel averages but it falls for 29 channel
average;  30\% fall in the signal suggests the signal   peaks 
within a velocity width of $570 \, \rm km \, sec^{-1}$ (19-channel average),
but there is detectable flux for velocity width $\simeq 1000 \, \rm km \, sec^{-1}$. (b)  the maximum signal-to-noise is obtained for the 
velocity width $\simeq 570 \, \rm km \, sec^{-1}$, (c) residual flux is 
seen on the wings of the coadded spectra in almost all cases, in particular
in the primary-beam corrected cases. 

It is conceivable that the presence of residual flux 
is owing to the width of the signal being wider than 
  $1000 \, \rm km \, sec^{-1}$, even though it is not 
supported by simulations (Khandai et~al 2011).  However, this 
hypothesis is less tenable because we do not see significant 
flux for the primary-beam uncorrected case, even though
the central enhancement is prominent.  A more likely 
hypothesis is that the residual flux in the line wings has other
origins: (a) presence of unsubtracted continuum, (b) systematics in spectral 
domain picked while correcting for the primary beam and possible edge effects.

The impact of edge effects is seen in Figure~1 and 
also  slightly discordant first and final
channel fluxes for the three different velocity-averaged  spectra 
(Figures~2--4). 
For 9-.  19-, and 29-channel averages we use channels: 10--90, 3--97,
7--93.  We   detect some edge effects which are seen
in particular in the 19-channel averaging which uses information 
from channels closer to the edge.

We consider the 29-channel averaged spectrum to estimate the level of 
continuum residual and systematics  in the coadded spectra. The 
 average flux of all the channels
except the central channel (average over $50\pm14$ channel) is: (0, 10,  6.7)~$\mu$Jy for the primary-beam uncorrected, corrected, and optimal case, respectively. The bootstrapping method redistributes  the line flux and also serves to 
remove systematics but it leaves the level of continuum unchanged. Therefore, 
the average of bootstrapped spectral points gives us an estimate of 
the level of residual continuum (Figure~3); this average is: (1.4, 4.4,  1.7)~$\mu$Jy for the primary-beam uncorrected, corrected, and optimal case, 
respectively. 

It follows  that the level of flux in the line wings is more than the residual
continuum. As already noted above, the primary-beam uncorrected
cube has relatively low levels of flux on the wings and the subtraction 
of residual continuum flux  removes the flux on the wings below
1$\sigma$ (Table~1). However, in the 
primary beam corrected uniform and optimal-weighing
cases the average flux on the wings is still significant, which 
tends to suggest detectable flux even in the line wings at 1.5--2$\sigma$
 levels. For instance, we compute the probability of the flux
level seen in the line wings in the highest signal-to-noise case: 
 19-channel averaged cube weighted  optimally (Figure~3) 
 by using bootstrapping method, we find the 
probability to be $\simeq 0.05$, roughly corresponding to a 2$\sigma$ 
detection.    However, the analysis
of the primary beam uncorrected  case clearly suggests  
that the line wing flux  is attributable to a mix of residual
continuum flux and systematics. The presence of systematics is also seen
in the asymmetrical shape of the lines.

To get a conservative estimate of the signal, we assume that
there is no flux in the cube for velocity widths exceeding the 
29-channel average, and  remove the flux in the line
wings estimated from the 
29-channel averaged spectra from all the channels in the 
corresponding cubes. 
The error on the central enhancement is obtained by the bootstrapping method (Table~1). Our final  results are shown in Figure~6 for all the cases 
and they can be summarized as:
\begin{itemize}
\item[1.] The signal-to-noise is maximum for the 19-channel averaging case
and  it  lies in the range $2.7\hbox{--}4.7$. 
\item[2.] The signal-to-noise is the lowest  for the 29-channel averaging case
and   is  in the range $1.8\hbox{--}3$.
\item[3.] As expected, the signal-to-noise is  maximum for
 the optimally-weighted case. 
\end{itemize}

\section{Interpretation and Conclusions} 
We can directly compare our results with the simulations of Khandai et~al. (2011). Their results showed that the stacked signal could have peak strength
varying between 10--30~$\mu$Jy  and 
velocity width in the range $500\hbox{--}1000 \, \rm km \, sec^{-1}$
 (Figure~5 of their paper).

Khandai et~al. (2011)  considered many  models, for 
the global HI content corresponding to $\Omega_{\rm HI} = 10^{-3}$,
 to represent the HI content of the DEEP2 galaxies at $z \simeq 1$, 
 based on different mass cuts 
on the dark matter haloes (e.g. Figure~7 of their paper). Our results
are   in  agreement with their model~2 and~3, corresponding
 to a  mass
cut-off  on the dark matter haloes in the range  $10^{12}\hbox{--}10^{12.5} \, \rm M_\odot$. 
For these  model, the line  flux peaks in the 
range  $\simeq 20\hbox{--}30 \, \rm \mu$Jy and falls by roughly a factor 
of 10 within a velocity width of $800\hbox{--}1000 \, \rm km \, sec^{-1}$ from
the line center.  This also 
means that the HI mass function at $z \simeq 1$ is more skewed towards the 
higher mass end as compared to the $z \simeq 0$ mass function (Zwaan et~al. 2005);  for detailed discussion, see Khandai et~al. 2011).

Uncertainties in the estimate of 
 HI content of the universe at $z \simeq 1$ partly
 arises from the fact that at  this redshift   the Lyman-$\alpha$ line is  
 not directly accessible to ground based telescopes. At this redshift,
the determination of the global HI content is enabled by HST observations, 
observation of associated MgII absorbers (Rao et~al. 2006), 
the detection of HI in absorption against bright radio sources (Kanekar et~al. 2009) and,
more recently, through the cross-correlation of optical galaxies with redshifted
HI emission (Chang et~al. 2010). All these 
methods yield the HI content of the universe at $z \simeq 1$, 
 $\Omega_{\rm HI}  \simeq 10^{-3}$ is roughly constant up to $z \simeq 4$ (Noterdaeme et~al. 2012, Rao et~al 2006).

Our result can potentially  be interpreted in two possible ways: either 
the value of 
$\Omega_{\rm HI}$ is higher than indicated by absorption studies or the 
DEEP2 galaxies provide a data set biased towards higher HI masses. The
 simulations of Pontzen et al.(2008) suggest that damped 
Lyman-$\alpha$ systems are housed in dark matter halos
of masses smaller than $10^{11} \, \rm M_\odot$; from the measurement
of mass-to-light ratio of DEEP2 galaxies, these galaxies are in dark matter
 halos almost an order of magnitude more massive,  (Khandai et~al. 2011, 
Figure~6).   Alternatively, 
the mass function of the HI halos is tilted towards higher 
masses as compared to low redshifts. As noted above, a comparison of 
our results with the simulations of Khandai et~al (2011),   suggests that
 our results can be interpreted as arising from 
a HI mass distribution skewed towards the higher mass end, without
invoking any enhancement in the global HI content. However, it 
is possible that the result could be partially owing 
to the uncertainty in the  HI content at $z \simeq 1.3$.

In conclusion: we report the first tentative detection of HI
 at $z \simeq 1.3$.  Our 
results are  based on GMRT observation  of  DEEP2 fields  and the coaddition
of the signals at the locations of these galaxies. If confirmed, this 
would be the highest redshift at which the HI has been detected in emission. 
These results can be understood  within the framework of existing 
data  on the global HI content at $z \simeq 1$  and 
detailed  modelling of HI at this redshift based on N-body  simulations. 

Currently, we are analysing  GMRT data for additional DEEP2 fields 
 and will  obtain more data in the coming GMRT cycle. We aim 
to put our results on a firmer footing by a detection at a  higher level of 
significance.

\begin{deluxetable}{ccccccc}
\tablenum{1}
\tablecolumns{8}
\tablewidth{0pc}
\tablecaption{HI Line  results}
\tablehead{
\colhead{}&\colhead{Uncorrected }& &\colhead{Corrected}   & & \colhead{Corrected} 
\\ 
\colhead{}& \colhead{}& &  (Uniform) & & (optimal) 
\\
\colhead{}&\colhead{($\mu$Jy)}& &\colhead{($\mu$Jy)} & & \colhead{($\mu$Jy)}

}
\startdata
{\bf 9-channel average} & & & & & & \nl
& Signal & Noise & Signal & Noise & Signal & Noise \nl
DEEP2 galaxies & 16.1 & 6.4 & 34.6 & 11.8 & 22.2 & 7.6 \nl
Random sight lines  & NA  & 4.4 & NA & NA & NA & NA \nl
Scrambled redshifts & NA  & 4.3  & NA  & 10 & NA & 4.1  \nl
 {\bf 19-channel average} & & & & & & \nl
& Signal & Noise & Signal & Noise & Signal & Noise \nl
DEEP2 galaxies & 13.2 & NA & 32.9 & NA & 21.6 & NA  \nl
Random sight lines  & NA  & 3.4 & NA & NA & NA & NA \nl
Scrambled redshifts & NA  & 3.8  & NA  & 8.6 & NA & 3.1  \nl
{\bf 29-channel average} & & & & & & \nl
& Signal & Noise & Signal & Noise & Signal & Noise \nl
DEEP2 galaxies & 9.2 & NA & 24  & NA & 15.7 & NA \nl
Random sight lines  & NA  & 2.8 & NA & NA & NA & NA \nl
Scrambled redshifts & NA  & 3.4  & NA  & 7.5 & NA & 2.8  \nl
\enddata
\tablecomments{The first three rows give the results based on 9 channel
averaging. The First column computes the average and the RMS from the 
coadded spectrum containing DEEP2 galaxies. The second row gives the noise
estimates based on the 40000 random lines of sight. The third row gives 
the error  when DEEP2 galaxy redshifts are scrambled. The other
rows present results for the other two velocity-averaged spectra} 
\end{deluxetable}

\begin{figure}
\epsfig{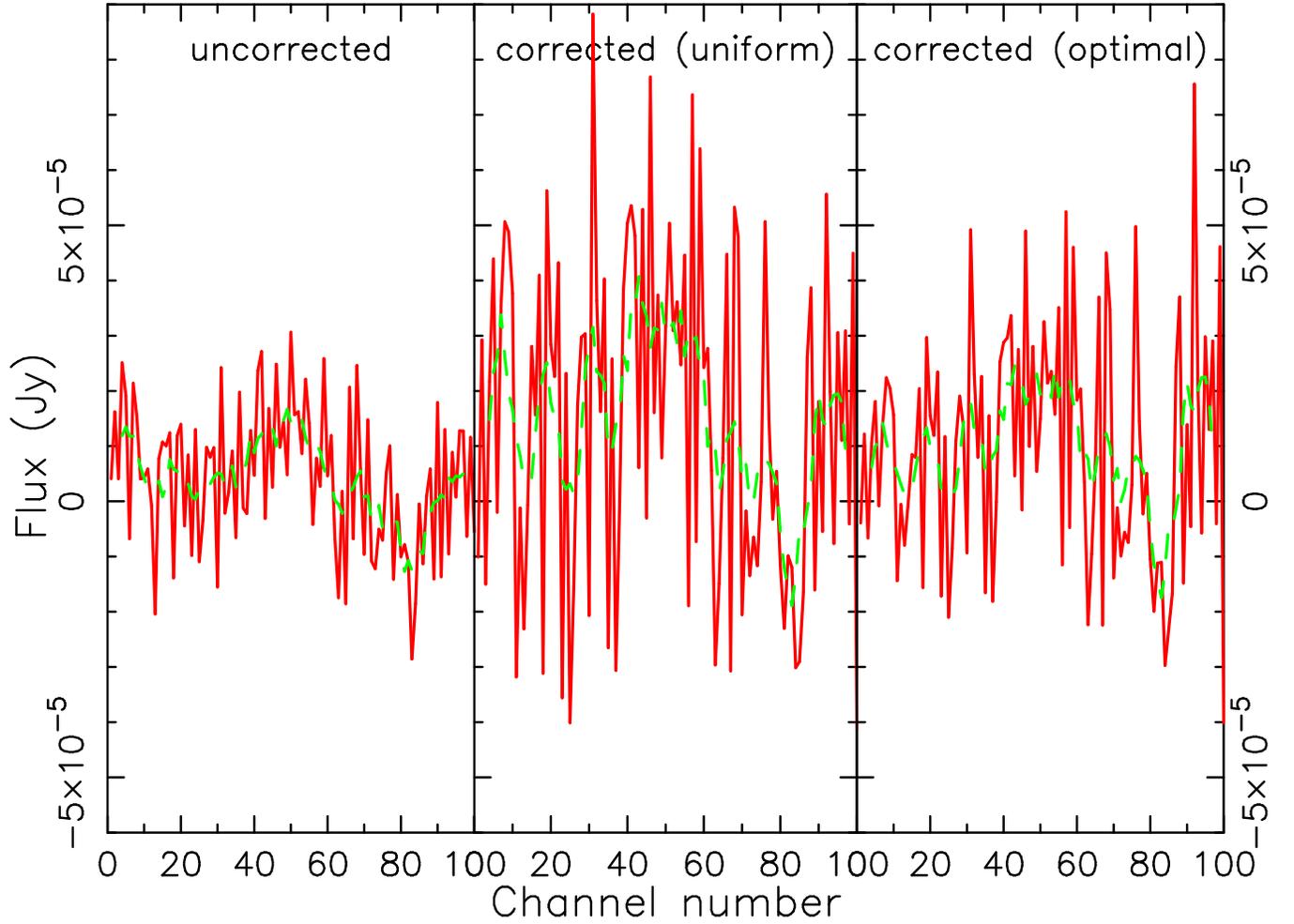}
\caption{ Co-added spectra toward DEEP2 galaxies.
Left panel: The spectra are shifted 
and coadded at the central channel (number 50). The 
coadded spectrum contain 489 sources. The two curves correspond to:
raw (solid, red), running average of  7 channels (dashed, green).
The channel width is $\simeq 30 \, \rm km s^{-1} $.  Middle and Right panels: 
The same as the Left panel  for primary beam corrected cube for uniform
and optimal weighing, respectively. There
are 371 sources in the coadded spectrum. 
}
\label{fig:f1}
\end{figure}

\begin{figure}
\epsfig{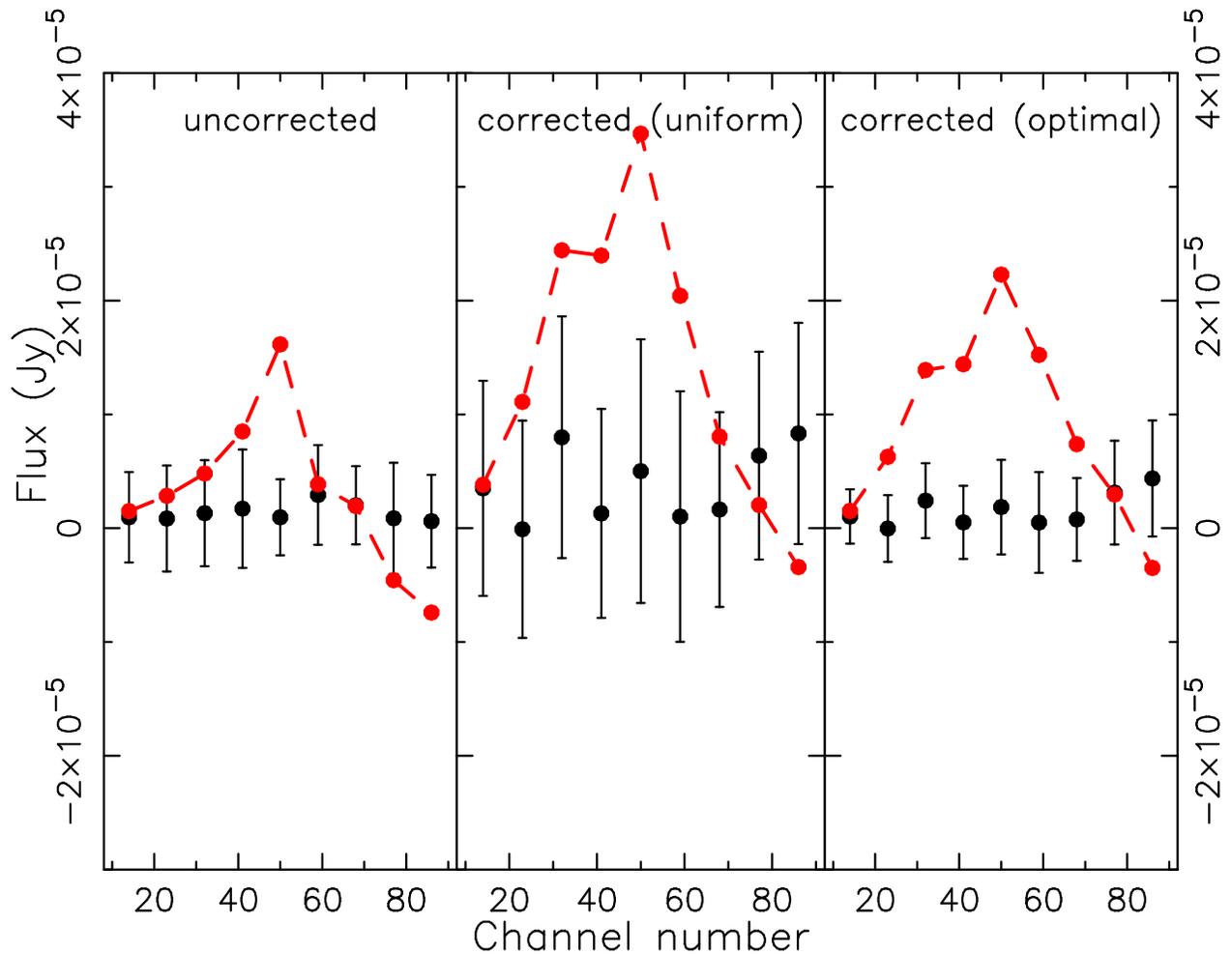}
\caption{Co-added HI signal toward the DEEP2 galaxies.
The red dashed line shows the signal obtained from coadding
and averaging over 9 channels.
The data points and errors shown are  obtained by computing the 
mean and the RMS of 20 realizations at each spectral 
point,  from realizations obtained by  scrambling  the redshifts of DEEP2 galaxies. 
The Left, center, and right panels correspond to primary beam uncorrected, corrected (uniform),
and corrected (optimal) cases, respectively
}
\label{fig:f5}
\end{figure}

\begin{figure}
\epsfig{file=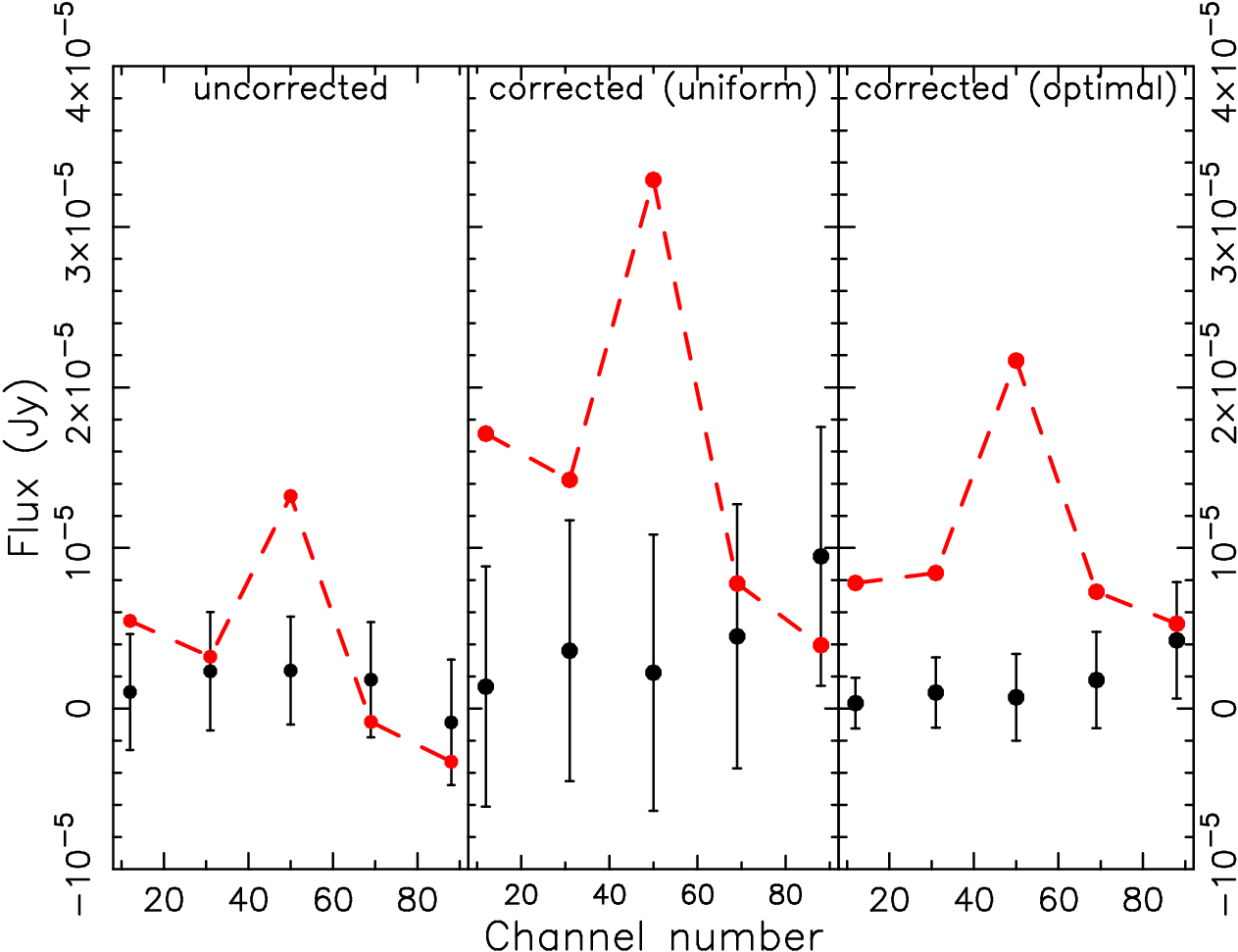,width=13cm,angle=270}
\caption{Co-added HI signal toward the DEEP2 galaxies.
The red dashed line shows the signal obtained from coadding
and averaging over 19 channels.
The data points and errors shown are  obtained by computing the 
mean and the RMS of 40 realizations at each spectral 
point,  from realizations obtained by  scrambling  the redshifts of DEEP2 galaxies.}
\label{fig:f5}
\end{figure}

\begin{figure}
\epsfig{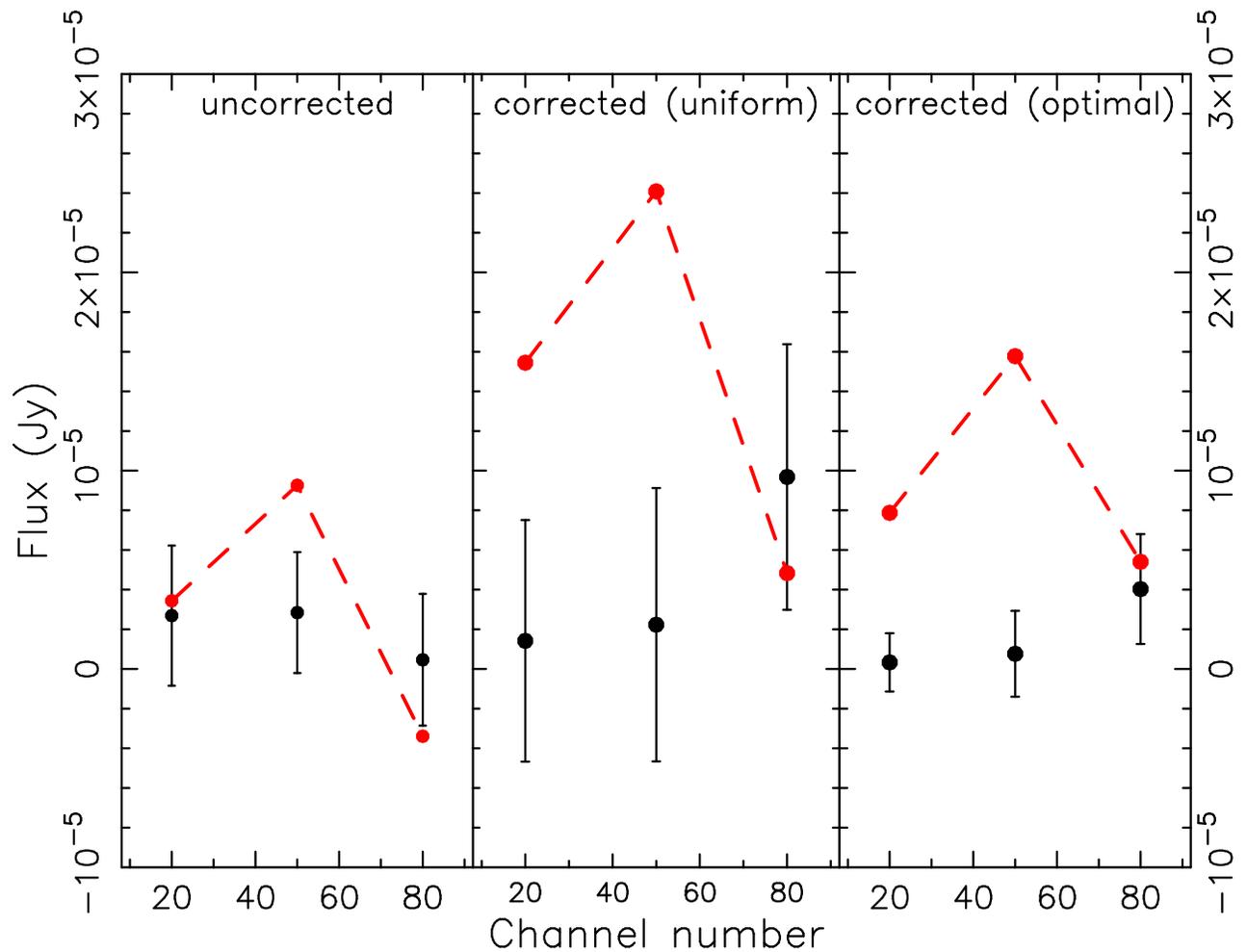}
\caption{Co-added HI signal toward the DEEP2 galaxies.
The red dashed line shows the signal obtained from coadding
and averaging over 29 channels.
The data points and errors shown are  obtained by computing the 
mean and the RMS of 60 realizations at each spectral 
point,  from realizations obtained by  scrambling  the redshifts of DEEP2 galaxies.}
\label{fig:f5}
\end{figure}

\begin{figure}
\epsfig{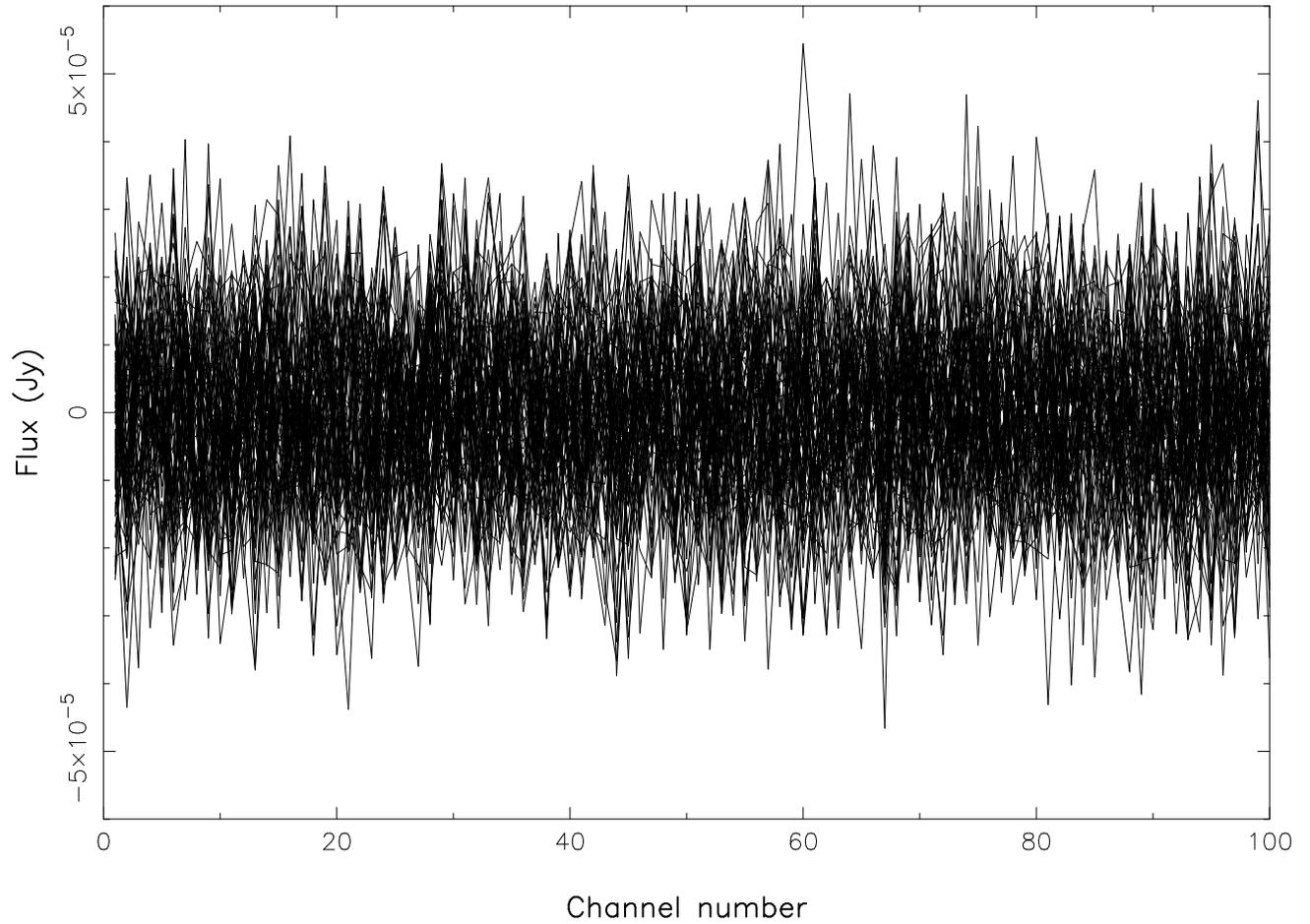}
\caption{ Noise in the cube toward blank positions.
Coadded spectra for 80 realizations (total 40000 sources, 
each realization contains the coadded spectra of 500 sources) obtained from
random lines of sights from  the primary beam uncorrected cube. 
}
\label{fig:f3}
\end{figure}

\begin{figure}
\epsfig{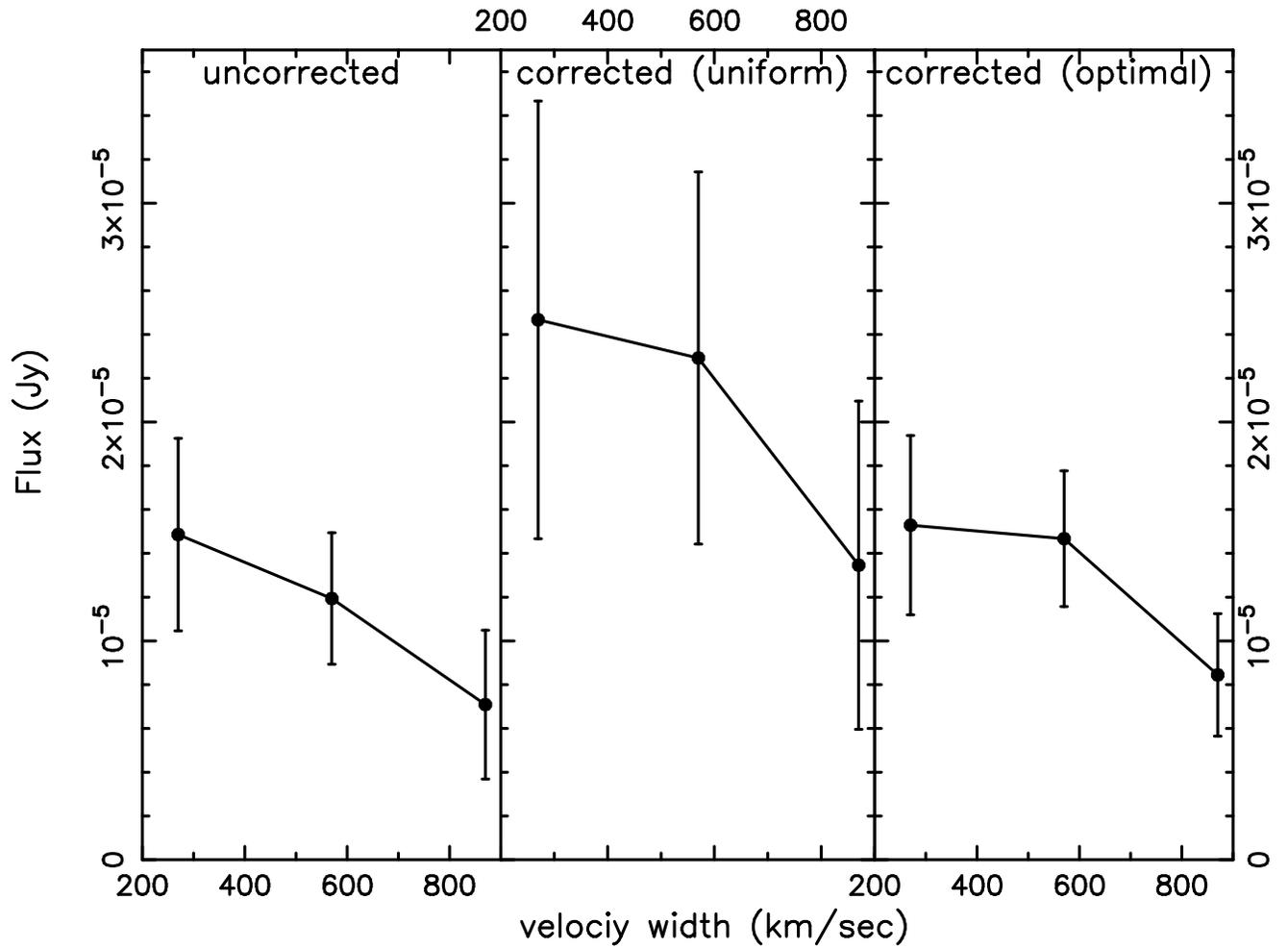}
\caption{The change in peak flux as a function of velocity
width is shown for the three data cubes. }
\label{fig:f6}
\end{figure}

\section*{Acknowledgement}
We thank Rajaram Nityananda for many 
useful suggestions and  the staff of the GMRT who have made these observations possible.
The GMRT is run by the National Centre for Radio Astrophysics of the Tata
Institute of Fundamental Research.

\end{document}